\documentclass[prl,aps,twocolumn,showpacs]{revtex4}
\usepackage{amsmath,graphicx,epsfig}

\def\prn#1{{\left(#1\right)}}

\def\sbrk#1{{\left[#1\right]}}
\def\abrk#1{{\langle#1\rangle}}

\def\ts#1{{_{\mbox{\scriptsize #1}}}}

\usepackage{euscript}

\def\sA{{\ensuremath{\EuScript A}}}

\def\sV{{\ensuremath{\EuScript V}}}

\def\fig_width{3. in} 
\draft

\newlength{\defbaselineskip}
\newcommand{\setlinespacing}[1]%
           {\setlength{\baselineskip}{#1 \defbaselineskip}}

\begin{document}

\title{Constraints on anomalous spin-spin interactions from spin-exchange collisions} 

\author{D. F. Jackson Kimball}
\email{derek.jacksonkimball@csueastbay.edu}
\affiliation{Department of Physics, California State University --
East Bay, Hayward, California 94542-3084, USA}

\author{Alec Boyd}
\affiliation{Physics and Astronomy Department, Pomona College, Claremont, CA 91711-6324, USA}

\author{D. Budker}
\affiliation{Department of Physics, University of
California at Berkeley, Berkeley, California 94720-7300} \affiliation{Nuclear Science
Division, Lawrence Berkeley National Laboratory, Berkeley, California 94720, USA}

\date{\today}



\begin{abstract}
Measured and calculated cross sections for spin-exchange between alkali atoms and noble gases (specifically sodium and helium) are used to constrain anomalous spin-dependent forces between nuclei at the atomic scale ($\sim 10^{-8}~{\rm cm}$).  Combined with existing stringent limits on anomalous short-range, spin-dependent couplings of the proton, the dimensionless coupling constant for a heretofore undiscovered axial vector interaction of the neutron arising from exchange of a boson of mass $\lesssim 100~{\rm eV}$ is constrained to be $g_A^n/\sqrt{4 \pi \hbar c} < 2 \times 10^{-3}$.  Constraints are established for a velocity- and spin-dependent interaction $\propto \prn{ \mathbf{I} \cdot \mathbf{v} } \prn{ \mathbf{K} \cdot \mathbf{v}  }$, where $\mathbf{I}$ and $\mathbf{K}$ are the nuclear spins of He and Na, respectively, and $\mathbf{v}$ is the relative velocity of the atoms.  Constraints on torsion gravity are also considered.
\end{abstract}








\maketitle

In the continuing effort to understand interactions between particles, the possibility of the existence of new fundamental forces often arises.  Here we investigate constraints on anomalous, heretofore undiscovered spin-spin interactions between atomic nuclei based on the comparison between experimental measurements and theoretical estimates of spin-exchange cross sections between alkali-metal atoms and noble gas atoms.

Recently, Dobrescu and Mocioiu \cite{Dob06} parameterized the long-range potentials induced by the exchange of new gauge bosons or spinless particles between fermions assuming rotational invariance, and reviewed existing limits on such potentials from measurements of macroscopic forces and astrophysical observations.  They noted that although there have been numerous searches for long-range spin-independent forces (see, for example, Ref.~\cite{Ade09} for a review), there are far fewer searches for spin-dependent forces, which in fact offer a richer variety of observables.  Among the experimental searches for spin-dependent forces, most have explicitly searched for monopole-dipole or dipole-dipole interactions, such as those that would arise due to axion exchange \cite{Moo84}.

Recent experiments have significantly improved constraints on anomalous dipole-dipole interactions between neutrons \cite{Gle08,Vas09} and monopole-dipole interactions between electrons and nucleons \cite{Hec08}. Previous experiments have constrained monopole-dipole interactions between nuclei \cite{Ven92} and anomalous dipole-dipole interactions between electrons \cite{Chu93,Bob91} and electrons and nuclei \cite{Win91}.  These experiments also place limits on exotic spin-dependent interactions predicted by a variety of novel theories \cite{Heh76,Sha02,Ham02,Kos08,Wei78,Dob05,Geo07,Ark05}.

The aforementioned experimental constraints on anomalous spin-dependent forces between fermions have all been obtained at the laboratory scale, where the distance between the interacting particles is on the order of a few centimeters or larger.  The only constraint on anomalous spin-dependent forces below the centimeter scale was obtained by Ramsey in 1979 \cite{Ram79} by comparing theoretical calculations and experimental measurements of the tensor force between protons in molecular hydrogen, which extends limits on anomalous spin-spin interactions between protons down to the molecular scale ($\sim 10^{-8}~{\rm cm}$).   Because of the possibility that new forces may have limited range, it is of considerable interest to find experimental techniques to search for anomalous spin-dependent interactions at short distances.

Constraints on anomalous spin-dependent interactions at the scale of interatomic distances can be obtained from measurements and calculations of spin-exchange collision cross sections.  In the following we investigate the possible contribution of anomalous spin-spin interactions to spin exchange between light alkali-metal atoms and noble gases, for which the spin-exchange cross sections are typically quite small ($\sim 10^{-24}~{\rm cm^2}$ \cite{Wal89,Wal97}).  The interaction of colliding atoms are dominated by the spin-independent interatomic potential $V(R)$, where $R$ is the internuclear separation, and spin exchange is induced by a much smaller, spin-dependent potential arising from electric and magnetic fields generated inside the atomic cores.  If the usual spin-dependent potentials arising from electromagnetic interactions are replaced with anomalous spin-dependent potentials generated by the exchange of new force-mediating particles, we can calculate the contribution of such anomalous potentials to the spin-exchange cross sections and thereby constrain the range and coupling of such hypothetical particles from the agreement between experiment and theory.

Experimental data on spin-exchange cross-sections between light alkali-metal atoms and noble gases are rather sparse; here we consider one of the better studied systems, collisions between Na and $^3$He, for which the binary spin-exchange cross-section $\sigma\ts{SE}$ was measured in Refs.~\cite{Sob72,Bor03}, and calculations have been performed in Refs.~\cite{Wal89,Tsc09} (see Table~\ref{Table:SE-cross-sections}).  In addition to providing the smallest cross sections, focusing on light atoms allows for the simplest calculations since the interaction is dominated by binary collisions of relatively short duration, minimizing the contribution of the spin-rotation interaction and three-body effects \cite{Wal89,Wal97}.

\begin{table}
\caption{Comparison of experimental measurements and theoretical calculations of the binary spin-exchange cross-section $\sigma\ts{SE}$ for Na and $^3$He collisions (scaled to temperature $T=100^\circ{\rm C}$ according to a $T^{-1}$ dependence, as in Ref.~\cite{Wal89}).  Uncertainties in theoretical estimates of cross-sections are from Refs.~\cite{WalPrivate10,TscPrivate10}.}
\medskip \begin{tabular}{ccc} \hline \hline
$\sigma\ts{SE}$ (Experiment) & $\sigma\ts{SE}$ (Theory) & Ref. \\
\hline
\rule{0ex}{3.6ex} ~~~$1.26(22) \times 10^{-24}~{\rm cm^2}$~~~ & ~ & \cite{Sob72} \\
\rule{0ex}{3.6ex} ~~~$0.95(09) \times 10^{-24}~{\rm cm^2}$~~~ & ~ & \cite{Bor03} \\
\rule{0ex}{3.6ex} ~ & ~~~$1.2(6) \times 10^{-24}~{\rm cm^2}$~~~ & \cite{Wal89} \\
\rule{0ex}{3.6ex} ~ & ~~~$1.4(4) \times 10^{-24}~{\rm cm^2}$~~~ & \cite{Tsc09} \\
\hline \hline
\end{tabular}
\label{Table:SE-cross-sections}
\end{table}

The weighted average of the experimental measurements of $\sigma\ts{SE}$ for Na-$^3$He collisions is $1.0(1) \times 10^{-24}~{\rm cm^2}$ and the weighted average of the calculations is $\sigma\ts{SE} = 1.3(4) \times 10^{-24}~{\rm cm^2}$.  The experiments and calculations are in agreement at the 1-$\sigma$ level. Reference~\cite{Tsc09} suggests that the theoretical uncertainty is dominated by uncertainties in the calculation of the Fermi contact hyperfine interaction between the electron spin of Na and the nuclear spin of $^3$He.  Spin-exchange between light alkali-metal and noble gas atoms is described by the Hamiltonian
\begin{align}
H\ts{IS} = \alpha(R) \mathbf{I} \cdot \mathbf{S}~, \label{Eq:magnetic-dipole-Hamiltonian}
\end{align}
which couples the noble gas nuclear spin $\mathbf{I}$ to the alkali-metal electronic spin $\mathbf{S}$.  The coupling strength $\alpha(R)$ of the magnetic-dipole interaction is determined by the Fermi contact interaction, which can be approximated as \cite{Her65}
\begin{align}
\alpha(R) = \frac{16 \pi}{3} \frac{\mu_B \mu_I}{I} \left| \psi(R) \right|^2~, \label{Eq:alpha}
\end{align}
where $\mu_B$ is the Bohr magneton, $\mu_I$ is the magnetic moment of the noble gas nucleus, and $\psi(R)$ is the wave function of the valence electron of the alkali atom evaluated at the position of the noble gas nucleus $R$.  Due to the large kinetic energy acquired by the valence electron in the core potential of the noble gas atom, the valence electron wave function of the alkali-metal atom is enhanced by a factor of $\eta$ at $R$ compared to the wave function in the absence of the noble gas \cite{Her65,Wal87,Tsc09}.  Because of the sensitive $R$-dependence of the Fermi contact interaction, atomic theory for the electron wave functions is likely not accurate enough to exactly reproduce the experimental results.  Nonetheless, because there are independent measurements confirming that the valence electron wave functions are reasonably accurate (for example, measurements of spin-exchange frequency shifts \cite{Bab05} and hyperfine pressure shifts \cite{Hap72,Wal97}), we conservatively estimate that the contribution to the spin-exchange cross section from anomalous spin-spin interactions must be less than twice the theoretical uncertainty,
\begin{align}
\sigma\ts{SE}({\rm anom}) < 0.8 \times 10^{-24}~{\rm cm^2}~,
\label{Eq:limit-on-sigmaSE}
\end{align}
which is, in fact, nearly the entire value of the measured cross section.

We consider the possible contribution to the Na-$^3$He spin-exchange cross section of three of the anomalous spin-spin potentials proposed by Dobrescu and Mocioiu \cite{Dob06} based on exchange of a single new boson of mass $m$ within a Lorentz-invariant quantum field theory.  We assume the interaction is between the $^3$He nuclear spin $\mathbf{I}$ and the Na nuclear spin $\mathbf{K}$, since limits on anomalous interactions between the electron spin and the $^3$He spin from spin-exchange collision measurements are expected to be extremely weak because of the dominance of the previously discussed uncertainties in the theoretical calculation of the Fermi contact interaction. According to the nuclear shell model \cite{Kli52}, Na-$^3$He spin-exchange predominantly probes anomalous interactions between the valence protons of the Na nucleus and the valence neutron of $^{3}$He.  (For $^{23}$Na, with $K=3/2$, according to the $jj$ coupling scheme the nuclear spin arises mainly from three valence protons above the closed shell in the $3d_{5/2}$ state \cite{Kli52}; for $^{3}$He, the nuclear spin is primarily due to that of the single neutron in the $1s_{1/2}$ state.)  It should be noted that the only existing atomic-scale limits on anomalous spin-spin interactions are between protons \cite{Ram79}, and the most sensitive laboratory-scale searches have probed neutron-neutron interactions \cite{Vas09,Gle08}, electron-electron interactions \cite{Chu93,Bob91}, or electron-nucleon interactions \cite{Win91}.  There is no theoretical expectation for the relationship between coupling constants for different particles, thus the analysis of the present work places limits on previously unconstrained anomalous potentials between neutrons and protons.

If the new boson is an axion-like pseudoscalar ($P$), there arises a dipole-dipole potential between $^3$He and Na \cite{Moo84,Dob06},
\begin{widetext}
\begin{align}
\sV_3(R) = \frac{g_P^n g_P^p}{4 \pi \hbar c} \frac{\hbar^3}{4 m_n m_p c} \sbrk{ \mathbf{I} \cdot \mathbf{K} \prn{ \frac{mc}{\hbar R^2} + \frac{1}{R^3} }  - \prn{ \mathbf{I} \cdot \hat{ \mathbf{R} } } \prn{ \mathbf{K} \cdot \hat{ \mathbf{R} } }  \prn{  \frac{m^2c^2}{\hbar^2 R} + \frac{3mc}{\hbar R^2} + \frac{3}{R^3} } } e^{ - mc R / \hbar }~,
\label{Eq:V3}
\end{align}
\end{widetext}
where $g_P^n g_P^p/\prn{ 4 \pi \hbar c }$ is the dimensionless pseudoscalar coupling constant between the neutron and proton and $m_n$ and $m_p$ are their respective masses.  (The subscripts $i$ for the anomalous potentials $\sV_i(R)$ are chosen to match the notation of Ref.~\cite{Dob06}.)  If the new boson is a spin-1 axial vector ($A$) particle, there arises a Yukawa potential
\begin{align}
\sV_2(R)  = \frac{g_A^n g_A^p}{4 \pi \hbar c}  \frac{\hbar c}{R} \mathbf{I} \cdot \mathbf{K} e^{ - mc R / \hbar } ~, \label{Eq:V2}
\end{align}
as well as a velocity-dependent spin-spin potential that can contribute to the spin-exchange cross section,
\begin{align}
\sV_8(R) = \frac{g_A^n g_A^p}{4 \pi \hbar c} \frac{\hbar c}{R} \prn{ \mathbf{I} \cdot \mathbf{v} } \prn{ \mathbf{K} \cdot \mathbf{v}  } e^{ - mc R / \hbar } ~,  \label{Eq:V8}
\end{align}
where $\mathbf{v}$ is the relative velocity between the colliding atoms. (A vector or axial vector interaction can also produce a $\sV_3(R)$ potential.)  It should be noted that there are also more general theoretical possibilities that generate the above potentials, as well as a variety of other spin-spin interactions \cite{Dob06}.

To calculate $\sigma\ts{SE}({\rm anom})$ between Na and $^3$He for the various anomalous potentials considered above, we follow the procedure outlined in Refs.~\cite{Wal89,Wal97} for calculation of spin-exchange cross sections, replacing the Fermi contact interaction Hamiltonian \eqref{Eq:magnetic-dipole-Hamiltonian} with the various anomalous potentials [Eqs.~\eqref{Eq:V3}-\eqref{Eq:V8}].  Our method of calculation was checked by reproducing the results in Refs.~\cite{Wal89,Wal97} for the usual spin-exchange cross sections and frequency shifts for a variety of light alkali-metal/noble-gas pairs.  Because the interaction time between the alkali-metal atom and noble gas atom during the collision is sufficiently brief ($\sim 10^{-12}~\rm{s}$) so that spin precession can be ignored, first-order time-dependent perturbation theory can be used, yielding, after averaging over the angular variables \cite{Wal89},
\begin{equation}
\begin{split}
&\sigma\ts{SE}({\rm anom}) = \\ &\frac{2\pi\mu}{\hbar^2k_BT} \int_0^\infty e^{-w}dw \int_0^\infty b~ db \left| \int_{r_0}^\infty \frac{\sA_i(R) dR}{\sqrt{ 1 - \frac{b^2}{R^2} - \frac{V(R)}{w k_B T} }} \right|^2 , \label{Eq:SE-anom-cross-section}
\end{split}
\end{equation}
where $\mu$ is the reduced mass, $w=\mu v^2/(2k_BT)$, $r_0$ is the classical turning point of the trajectory, $b$ is the impact parameter, $V(R)$ is the Na-He interatomic potential curve, and $\sA_i(R)$ are the corresponding anomalous spin-exchange potential coefficients given by the average over possible directions of $\hat{\mathbf{R}}$ and $\hat{\mathbf{v}}$ for colliding atoms [analogous to Eq.~\eqref{Eq:alpha}],
\begin{align}
\sV_i(R) = \sA_i(R) \mathbf{I}\cdot\mathbf{K}~,
\end{align}
where
\begin{align}
\sA_2(R) & = \frac{g_A^n g_A^p}{4 \pi \hbar c}  \frac{\hbar c}{R} e^{ - mc R / \hbar }~, \label{Eq:sA2} \\
\sA_3(R) & = \frac{g_P^n g_P^p}{4 \pi \hbar c} \frac{\hbar^3 e^{ - mc R / \hbar }}{4 m_n m_p c} \prn{ \frac{2mc}{3\hbar R^2} - \frac{m^2c^2}{\hbar^2 R}}~, \label{Eq:sA3} \\
\sA_8(R) & = \frac{g_A^n g_A^p}{4 \pi \hbar c} \frac{\hbar c v^2}{3R} e^{ - mc R / \hbar } \nonumber
\\
& = \frac{g_A^n g_A^p}{4 \pi \hbar c} \frac{2\hbar c k_B T w}{3\mu R} e^{ - mc R / \hbar }~. \label{Eq:sA8}
\end{align}
\begin{widetext}
As noted, in the above calculation for $\sA_3(R)$ we carry out an average of $\prn{ \mathbf{I} \cdot \hat{ \mathbf{R} } } \prn{ \mathbf{K} \cdot \hat{ \mathbf{R} } }$ over all possible orientations of $\hat{\mathbf{R}}$ with respect to the spins, and an identical angular average over possible orientations of $\mathbf{v}$ is required for calculation of $\sA_8(R)$.  Expressing $\hat{\mathbf{R}}$ in terms of spherical coordinates centered at the alkali nucleus, $\hat{\mathbf{R}} = \prn{\sin\theta\cos\phi, \sin\theta\sin\phi, \cos\theta}$, we can write
\begin{align}
\prn{ \mathbf{I} \cdot \hat{ \mathbf{R} } } \prn{ \mathbf{K} \cdot \hat{ \mathbf{R} } } =  \mathbf{I} \cdot \prn{ \hat{ \mathbf{R} } \hat{ \mathbf{R} } } \cdot \mathbf{K}~,
\end{align}
where
\begin{align}
\hat{ \mathbf{R} } \hat{ \mathbf{R} }  = \left(
\begin{array}{ccc}
 \sin ^2(\theta ) \cos ^2(\phi ) & \sin ^2(\theta ) \sin (\phi ) \cos
   (\phi ) & \sin (\theta ) \cos (\theta ) \cos (\phi ) \\
 \sin ^2(\theta ) \sin (\phi ) \cos (\phi ) & \sin ^2(\theta ) \sin
   ^2(\phi ) & \sin (\theta ) \cos (\theta ) \sin (\phi ) \\
 \sin (\theta ) \cos (\theta ) \cos (\phi ) & \sin (\theta ) \cos
   (\theta ) \sin (\phi ) & \cos ^2(\theta )
\end{array}
\right)~.
\end{align}
\end{widetext}
The angular average can then be readily obtained:
\begin{align}
\frac{1}{4\pi}\int_0^\pi \int_0^{2\pi} \hat{ \mathbf{R} } \hat{ \mathbf{R} } \sin\theta d\theta d\phi = \frac{1}{3}\left(
\begin{array}{ccc}
 1 & 0 & 0 \\
 0 & 1 & 0 \\
 0 & 0 & 1
\end{array}
\right)~,
\end{align}
which yields the intuitive result
\begin{align}
\abrk{ \prn{ \mathbf{I} \cdot \hat{\mathbf{R}} } \prn{ \mathbf{K} \cdot \hat{\mathbf{R}}  } } = \frac{1}{3} \mathbf{I}\cdot\mathbf{K}~,
\end{align}
where $\abrk{ \cdots }$ indicates the average over different orientations of $\hat{\mathbf{R}}$, and similarly
\begin{align}
\abrk{ \prn{ \mathbf{I} \cdot \mathbf{v} } \prn{ \mathbf{K} \cdot \mathbf{v}  } } = \frac{v^2}{3} \mathbf{I}\cdot\mathbf{K}~.
\end{align}
A more detailed analysis of this problem using the Wigner $D$-functions is described in Ref.~\cite{Wal98}.  It should be noted that because of the angular averaging $\sA_3(R) \rightarrow 0$ as $m \rightarrow 0$.

For the Na-He interatomic potential curve $V(R)$, we use a fit to the experimental results obtained by Havey {\it{et al}}. \cite{Hav80} (coefficients are in atomic units, accuracy assumed at the level of the last significant digit),
\begin{align}
V(R) = 0.27816 \times e^{-0.799 R}~,
\end{align}
which offers a simple analytical form and is in good agreement with theoretical calculations (for example, Refs.~\cite{Kra71,Han79,Pas83,Par01}).

\begin{figure}
\center
\includegraphics[width=3.35 in]{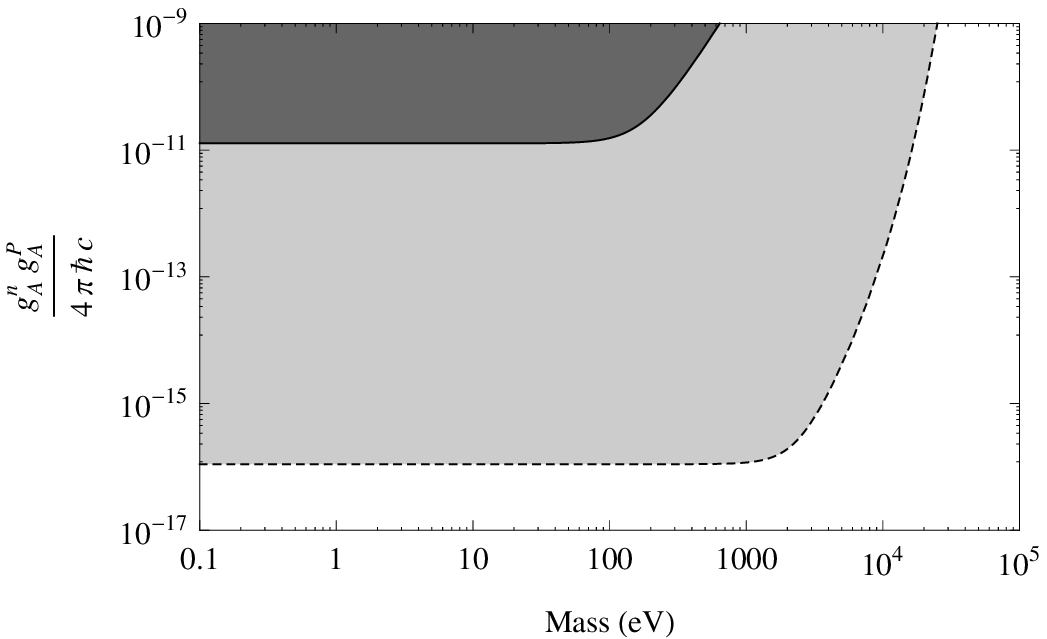}
\includegraphics[width=3.35 in]{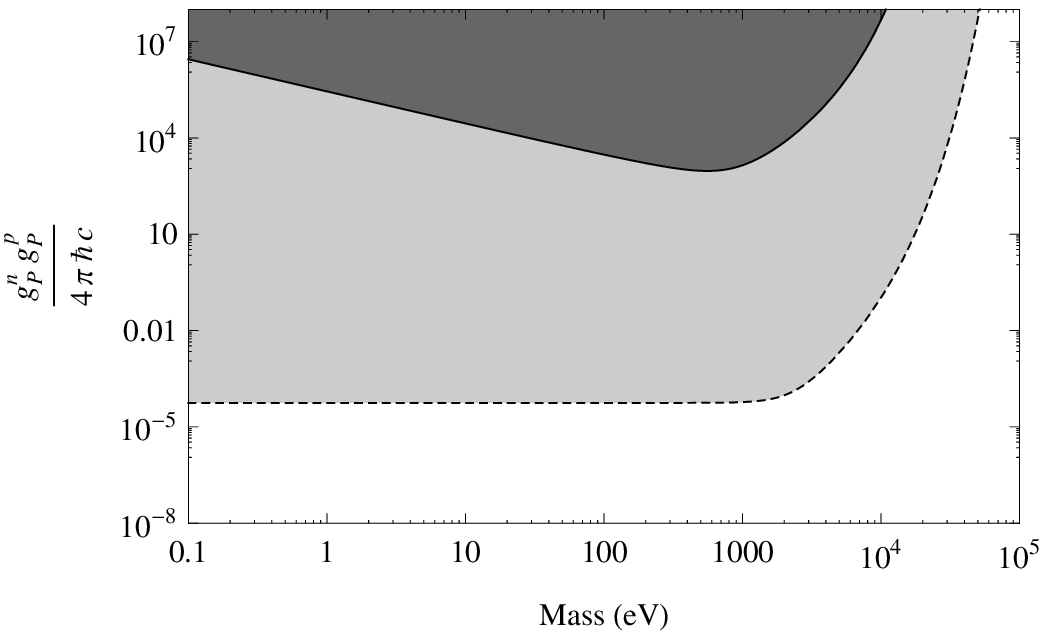}
\caption{Constraints on the dimensionless coupling constants $g_A^n g_A^p/\prn{ 4 \pi \hbar c }$ (upper plot) and $g_P^n g_P^p/\prn{ 4 \pi \hbar c }$ (lower plot) as a function of boson mass from calculation of the possible contribution of an axial vector (A) or pseudoscalar (P) interaction [$\sV_2(R)$ and $\sV_3(R)$, see Eqs.~\eqref{Eq:V2} and \eqref{Eq:V3}, respectively] to the Na-He spin-exchange cross section (solid line and dark gray fill).  The dashed line and light gray fill shows the corresponding constraints on $\prn{g_A^p}^2/\prn{ 4 \pi \hbar c }$ and $\prn{g_P^p}^2/\prn{ 4 \pi \hbar c }$ from calculations and measurements of molecular hydrogen spectra \cite{Ram79}. Note that the constraints roll off at a boson mass of $m \gtrsim 1000$~eV which corresponds to a Compton wavelength less than or equal to the typical internuclear separation in molecules.}\label{Fig:SElimits-spinspin-V2V3}
\end{figure}

\begin{figure}
\center
\includegraphics[width=3.35 in]{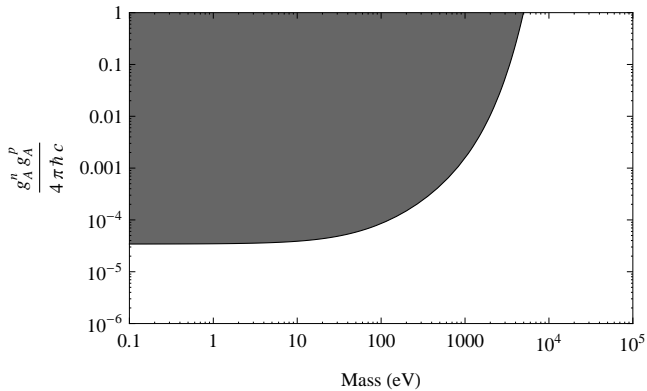}
\caption{Constraint on the dimensionless coupling constant $g_A^n g_A^p/\prn{ 4 \pi \hbar c }$ as a function of boson mass for the velocity- and spin-dependent-interaction $\sV_8(R)$ [Eq.~\eqref{Eq:V8}], based on Na-He spin-exchange cross section measurements and calculations. }\label{Fig:SElimits-spinspin-V8}
\end{figure}

We numerically integrate Eq.~\eqref{Eq:SE-anom-cross-section} for given values of $m$ and then limit the value of the dimensionless coupling constants based on the agreement between measured and calculated spin-exchange cross sections [Eq.~\eqref{Eq:limit-on-sigmaSE}].  The results of our calculations are shown in Figs.~\ref{Fig:SElimits-spinspin-V2V3} and \ref{Fig:SElimits-spinspin-V8}, which also show constraints on proton-proton interactions set by Ramsey \cite{Ram79}.
Note that the constraints on axial vector coupling constants are (upper plot, Fig.~\ref{Fig:SElimits-spinspin-V2V3}) are orders-of-magnitude more stringent than the constraints on pseudoscalar coupling constants (lower plot, Fig.~\ref{Fig:SElimits-spinspin-V2V3}).  This arises simply from the different forms of the anomalous potentials $\sA_2(R)$ and $\sA_3(R)$ [Eqs.~\eqref{Eq:sA2} and \eqref{Eq:sA3}, respectively].  At the typical minimum internuclear separation in collisions or molecules ($R \sim 0.1~{\rm nm}$), for equal values of the axial vector and pseudoscalar coupling constants, $\sA_2(R)$ is roughly twelve orders of magnitude larger than $\sA_3(R)$, explaining the large difference between the limits.

Our work establishes the first limits on the velocity- and spin-dependent interaction $\sV_8(R)$ [Eq.~\eqref{Eq:V8}, Fig.~\ref{Fig:SElimits-spinspin-V8}], with a long-range limit (boson mass $\lesssim 1~{\rm eV}$)
\begin{align}
\frac{g_A^n g_A^p}{4 \pi \hbar c} < 3.4 \times 10^{-5}~.
\end{align}

Limits on $\sV_2(R)$ and $\sV_3(R)$ from the work of Ramsey \cite{Ram79} on spectroscopy of molecular hydrogen are more stringent by several orders of magnitude, and can be used to limit the proton coupling constants.  For example, for a boson of mass $100~{\rm eV}$ (range $\lambda \approx 2 \times 10^{-7}~{\rm m}$),
\begin{align}
\frac{g_A^p}{\sqrt{4 \pi \hbar c}} < 10^{-8} & & \frac{g_P^p}{\sqrt{4 \pi \hbar c}} < 7.4 \times 10^{-3}~,
\end{align}
which when combined with our results provide the first limits on the neutron couplings at the atomic scale:
\begin{align}
\frac{g_A^n}{\sqrt{4 \pi \hbar c}} < 2 \times 10^{-3} & & \frac{g_P^n}{\sqrt{4 \pi \hbar c}} < 4 \times 10^{5}~.
\end{align}

\begin{figure}
\center
\includegraphics[width=3.35 in]{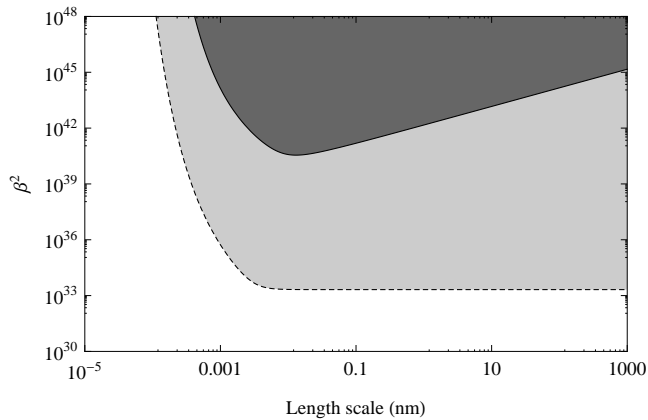}
\caption{Constraint on the dimensionless torsion gravity coupling constant $\beta^2$ as a function of interaction range $\lambda$ for neutron-proton interactions (solid line, dark gray fill --- from Na-He spin-exchange collision cross section measurements) and for proton-proton interactions (dashed line, light gray fill --- from molecular hydrogen spectroscopy \cite{Ram79}).}\label{Fig:torsion-gravity-limits}
\end{figure}

The above analysis can also be used to place bounds on torsion gravity.  According to general relativity, a purely tensor theory, the intrinsic spin of a particle is unaffected by the local gravitational field \cite{Kob62,Lei64,Heh90,Khr98,Sil05}.  However, in extensions of general relativity based on a Riemann-Cartan spacetime instead of a Riemann geometry, the gravitational interaction is described by a torsion tensor which can generate spin-mass and spin-spin interactions \cite{Heh76,Sha02,Ham02,Kos08}.  Depending on the specific model, these spin-dependent couplings can be long- or short-range \cite{Nev80,Nev82,Car94,Ham95}.  The spin-spin interaction generated by a standard propagating torsion field is similar to the dipole-dipole coupling $\sV_3(R)$ \cite{Ham95,Ade09}, and can be parameterized in terms of a dimensionless parameter $\beta$ and a length scale $\lambda$ [equivalent to the Compton wavelength for a particle of mass $m$ in Eq.~\eqref{Eq:V3}], where
\begin{align}
\beta^2 = \prn{ \frac{g_P^n g_P^p}{4 \pi \hbar c} } \times \prn{ \frac{2}{9} \frac{\hbar c}{G m_n m_p} }
\end{align}
where $G$ is the gravitational constant.  Constraints on short-range proton-proton and neutron-proton torsion gravity couplings are shown in Fig.~\ref{Fig:torsion-gravity-limits}.   The minimally coupled Dirac equation for a spin-$1/2$ particle \cite{Nev80,Nev82,Car94,Ham95} leads to $\beta = 1$ and $\lambda \rightarrow \infty$ \cite{Ham95}.  The recent experiment of Romalis and co-workers \cite{Vas09} sets the limit $\beta^2 < 2.2 \times 10^{28}$ for $\lambda \gtrsim 50~{\rm cm}$.

In comparison to the earlier work of Ramsey \cite{Ram79} on anomalous spin-dependent couplings of the proton, our limits on Yukawa [$\sV_2(R)$, Eq.~\eqref{Eq:V2}] and dipole-dipole [$\sV_3(R)$, Eq.~\eqref{Eq:V3}] couplings are orders-of-magnitude less stringent, emphasizing the superiority of molecular spectroscopy as a method for constraining anomalous spin-dependent forces at atomic scales.  In the future we plan to explore the possibility of using new molecular spectroscopy experiments to search for anomalous short-range spin-dependent interactions.

Also of interest is the recent work of Walker, Nelson, and Kadlecek \cite{Wal10} which demonstrates a method by which isotropic and anisotropic spin-exchange rates can be differentiated purely through experimental data.  Such a method could be adapted to enable direct experimental searches for short-range spin-dependent interactions using measurements of spin-exchange cross-sections without reference to theoretical calculations.

In conclusion, by comparing measurements \cite{Sob72,Bor03} and calculations \cite{Wal89,Tsc09} of the Na-$^3$He spin-exchange cross section $\sigma\ts{SE}$, we have determined a conservative upper limit to the possible contribution of anomalous spin-dependent interactions to $\sigma\ts{SE}$.  Employing a calculation modeled on the work of Walker \cite{Wal89,Wal97}, we have constrained three possible spin-dependent potentials induced by the exchange of new gauge bosons or spinless particles between fermions \cite{Dob06}.  We have established the first constraint on a velocity- and spin-dependent potential $\sV_8(R) \propto \prn{ \mathbf{I} \cdot \mathbf{v} } \prn{ \mathbf{K} \cdot \mathbf{v}  }$, where $\mathbf{I}$ and $\mathbf{K}$ are the nuclear spins of He and Na, respectively, and $\mathbf{v}$ is the relative velocity of the atoms.  We have also established the first constraints on anomalous spin-dependent neutron couplings at the atomic scale.  Our analysis also provides constraints on short-range torsion gravity fields.

\acknowledgments

The authors are grateful for enlightening conversations with Thad Walker, Micah Ledbetter, and Timur Tscherbul.  A.B. was an undergraduate research fellow at California State University - East Bay while working on this project.  This work has been supported by grants PHY-0652824 and PHY-0969666 from the National Science Foundation and Faculty Support Grants from California State University - East Bay.

\end{document}